\documentclass[twocolumn,pra,amsmath,amssymb,longbibliography]{revtex4-1}

\usepackage[hidelinks]{hyperref}
\usepackage{xcolor}
\usepackage{mathtools}
\usepackage{amsfonts}
\usepackage{braket}
\usepackage{braket}
\usepackage{ulem}

\newcommand{\tr}[1]{Tr[#1]}
\newcommand{\be}{\begin{equation}}
\newcommand{\ee}{\end{equation}}
\newcommand{\prj}[1]{|#1\rangle\langle#1|}
\newcommand{\matel}[3]{\langle#1|#2|#3\rangle}
\newcommand{\DDi}{\tilde{D}_i}
\newcommand{\DDj}{\tilde{D}_j}
\newcommand{\DDk}{\tilde{D}_k}

\renewcommand{\eqref}[1]{Eq.(\ref{#1})}
\newcommand{\figref}[1]{Fig.(\ref{#1})}
\newcommand{\tableref}[1]{Tab.(\ref{#1})}
\newcommand{\secref}[1]{Sec.\ref{#1}}
\newcommand{\appref}[1]{App.\ref{#1}}

\begin{document}

\title{Characterisation of multi-level quantum coherence without ideal measurements}
\date{\today}
\author{Benjamin Dive$^{*,1,2}$, Nikolaos Koukoulekidis$^{*,1}$, Stefanos Mousafeiris$^1$, and Florian Mintert$^1$}
\affiliation{$^*$These two authors contributed equally to this work}
\affiliation{$^1$Department of Physics, Imperial College, London SW7 2AZ, UK}
\affiliation{$^2$Institute of Quantum Optics and Quantum Information, Austrian Academy of Sciences, Vienna 1090, Austria}

\begin{abstract}
Coherent superpositions are one of the hallmarks of quantum mechanics and are vital for any quantum mechanical device to outperform the classically achievable. Generically, superpositions are verified in interference experiments, but despite their longstanding central role we know very little about how to extract the number of coherently superposed amplitudes from a general interference pattern. A fundamental issue is that performing a phase-sensitive measurement is as challenging as creating a coherent superposition, so that assuming a perfectly implemented measurement for verification of quantum coherence is hard to justify. In order to overcome this issue, we construct a coherence certifier derived from simple statistical properties of an interference pattern, such that any imperfection in the measurement can never over-estimate the number of coherently superposed amplitudes. We numerically test how robust this measure is to under-estimating the coherence in the case of imperfect state preparation or measurement, and find it to be very resilient in both cases.
\end{abstract}

\maketitle

%%%%%%%%%%%%%%%%%%%%%%%%%%%%%%
\section{Introduction}

The superposition principle allows wave mechanics, in particular quantum mechanics, to feature dynamics that are unthinkable for classical particles. The prospect of exploiting quantum coherence for applications in quantum computation, communication, metrology, and thermodynamics \cite{ref:Stahlke, ref:Knill, Zhang2017a, ref:Lostaglio, ref:Korzekwa} has resulted in numerous activities towards the classification and quantification of quantum coherence \cite{1367-2630-16-3-033007, Baumgratz2014a, Girolami2014, VonPrillwitz2015, ref:Winter, ref:Marvian, ref:Streltsov2, Streltsov2017}.

Those developments are inspired by earlier work in the theory of entanglement. There is, however, a central difference between entanglement and coherence that poses a fundamental challenge in its experimental characterisation. To create entanglement it is necessary to use coherent interactions between particles that go beyond Local Operations and Classical Communications (LOCC). It can however be detected using only local measurements and classical processing of the resulting data, e.g., in terms of Bell inequalities, witnesses or state tomography \cite{Horodecki2009, Friis2018}. Thus, verifying entanglement requires less challenging experimental tools than to prepare it.

This distinction between resources needed for preparation and detection does not typically exist for coherence. Coherence is always defined with respect to a basis and this is generically the only basis in which measurements can be performed. Creating coherence requires an operation that maps a basis state into a coherent superposition of basis states; detecting coherence requires a measurement in such a superposition basis. As the latter typically cannot be done, it is instead replaced with an operation that maps the state back to an incoherent one (essentially the reverse of the preparation step), followed by a projection onto one of the basis states. This results in the awkward situation that any measurement that is supposed to verify the successful preparation of a coherent superposition is reliable only under the assumption that coherent superpositions can be created.

As we show here, this is not an insurmountable obstacle. We can find suitable figures of merit that offer a detailed characterisation of coherence properties, but that do not require any assumption on the ability to realise operations that can create coherent superpositions.

Doing this first requires a rigorous definition of the aspects of coherence that we want to certify. For any given reference basis $\{\ket{j}\}$, one can define pure states $\ket{\psi}=\sum_j\psi_j\ket{j}$ with at least $k$ non-vanishing amplitudes $\psi_j$ to be $k$-coherent. Extending this, a mixed state $\rho$ is $k$-coherent if all decompositions $\rho=\sum_ip_i\prj{\psi_i}$ into pure states $\ket{\psi_i}$ with $p_i\ge 0$ contains at least one $k$-coherent pure state \cite{1367-2630-16-3-033007}. We denote the set of $k$-coherence state for a given Hilbert space by $C_k$, and have the natural relation that $C_{k+1} \subset C_k $ for $k\ge1$ and where $C_1$ is the full state space.

Following this definition, the concept of $k$-coherence is closely analogous to genuine $k$-partite entanglement. Most of the prior literature on quantum coherence has not yet addressed this fine classification of different classes of coherence, but there are figures of merit that characterize $k$-coherence quantitatively \cite{1367-2630-16-3-033007,Ringbauer2018} or qualitatively \cite{VonPrillwitz2015}. Almost all existing approaches do rely on the assumption that measurements can be performed reliably in a basis other than that of the $1$-coherent states, which is highly problematic for the reasons described above. The only exception we are aware of is from one of the proposals in \cite{Ringbauer2018} where the coherence is instead bounded by the probability of success of a quantum game, which comes with its own assumptions about the dynamics on the system and the measurements performed. Our method does away with these different assumptions and instead requires only the acquisition of relative phases and the ability to perform some rank-$1$ measurement afterwards.

We envision an experiment similar to the famous Ramsey sequence. This involves a preparation unitary $U_p$ such that $U_p\ket{0}=\sum_j\psi_j\ket{j}=\ket{\psi}$, followed by an evolution $U(t)$ generated by the system Hamiltonian $H$ for a time $t$.
This is followed by an effective projection onto a state $\ket{\chi}= \sum_j \chi_j \ket{j}$ which is realised by the unitary evolution $U_r$, defined by $U_r^\dagger\ket{0} =\ket{\chi}$, and a subsequent projection onto the basis state $\ket{0}$. As such, the probability of getting a `click' in the detector for an initial pure state $\ket{0}$ is given by $p(t)=|\matel{\chi}{U(t)}{\psi}|^2$. This defines the interference pattern that is observed.

The coherence of $\ket{\psi}$, with respect to the eigenbasis of $H$, can be characterised in terms of the statistical moments of this probability distribution, $M_q=\langle p^q\rangle$, where the average is taken over the period of the dynamics. When $\ket{\chi}$ is promised to be an equal superposition of all the eigenstates of $H$ (a state denoted by $\ket{W}$), these moments provide a rigorous indicator of $k$-coherence. That is, there is a threshold value such that moments above this threshold value can only be achieved with states that are at least $k$-coherent \cite{VonPrillwitz2015}. The intuition behind this is that the interference pattern of higher coherent states exhibit higher peaks and deeper troughs than low coherent states; in an analogous way to how the interference pattern of a diffraction grating and a double slit differ. This behaviour can be detected with the statistical moments, with higher moments being more sensitive to the more extreme peaks and troughs.

As argued above, it is highly problematic to assume that the desired projection onto the state $\ket{W}$ can be performed reliably. Assuming that such a projection was performed when a different measurement was realised can suggest a higher degree of coherence than there is. This can easily be seen with the extreme case of $\ket{\chi} = \ket{0}$. In this case $p(t)$ is maximised with the incoherent initial state $\ket{0}$, and since this holds for all $t$, also all moments adopt their maximum value for this state. Erroneously implementing a measurement including the projection onto the state $\ket{0}$ rather than the projection onto a balanced superposition of all basis states is certainly not a realistic experimental scenario, but it helps to illustrate that uncontrollable experimental imperfections can result in wrong conclusions if assumptions on the type of measurement are made. In order to have trusted certification, we require a function that can identify coherence in the case of suitable measurements, but that does not result in false positives.

In this paper we introduce a family of functions which do this, based on the ratio of moments of an interference pattern. We will show that those are convex functions of a quantum state, which makes them directly applicable to mixed states. The maximum value that such functions can adopt for a $k$-coherent state will be shown to be bounded from above independently of the Hamiltonian $H$ and the projector $\ket{\chi}\bra{\chi}$. Experimental limitations in the realization of the desired measurement will thus not result in wrong conclusions on the coherence properties of the state, but will in the worst case only result in the failure to exceed the threshold.

The construction of these coherence certifiers is presented in \secref{sec:math}, where their properties are also discussed. The technical aspects of the proofs are left to the appendices. In the cases where the exact threshold values are not known, we use numerical methods to approximate them; a discussion of these results is given in \secref{sec:num_threshold}. This is followed in \secref{sec:exp} by a discussion of the ability of the proposed framework to verify $k$-coherence in the presence of various imperfections, and we conclude in \secref{sec:conclusion}.

%%%%%%%%%%%%%%%%%%%%%%%%%%%%%%
\section{Coherence certifier}
\label{sec:math}
To talk in precise terms about the coherence certifiers we introduce, it is necessary to specify exactly the range of systems under consideration. The coherence of a state is defined with respect to a basis, and the natural basis to use for a Ramsey-like experiment is the eigenbasis of the system Hamiltonian. We make no restrictions on this Hamiltonian other than it being time-independent and having a discrete and commensurate spectrum (all finite Hamiltonians are discrete and $\epsilon$-close to being commensurate). It may contain some degeneracies but, as degenerate levels always have the same relative phases, these will never get picked up by the interference pattern and so the amount of coherence would be underestimated. As we are only lower bounding the coherence, this is not a problem.
In order to simplify the analysis it is therefore convenient to ignore these degeneracies and, furthermore, expand the Hilbert space of the system by adding new levels such that the spectrum of the Hamiltonian is equally spaced. As this does not affect the evolution of the physical state, there is no loss of generality in only considering Hamiltonians
\begin{align}
H = \sum_{n} n \ket{n}\bra{n}\ ,
\label{eq:BasicHamiltonian}
\end{align}
with the spectrum of a harmonic oscillator. For the certifier of coherence we introduce below, any anharmonicity in the physical Hamiltonian will lead to less coherence being measured, and therefore cannot result in a false certification of the amount of coherence present in the state.

As discussed in the introduction, the basic objects we use to study coherence are the moments of the interference pattern. The $n$\textsuperscript{th} moment is
\begin{align}
\label{eq:defineMoment}
M_n(\rho, \ket{\chi}) &= \frac{1}{2\pi} \int_0^{2\pi} p(t)^n \,dt \\
&= \frac{1}{2\pi} \int_0^{2\pi} \braket{\chi | e^{-i H t} \,\rho\, e^{i H t} | \chi}^n \,dt, \nonumber
\end{align}
where the duration of the integral is due to the energy scale picked in \eqref{eq:BasicHamiltonian}.
The key object of interest is the ratio
\be
R_n(\rho, \ket{\chi}) = \frac{M_n}{M_1^{n-1}}
\ee
of the moments $M_n$ and $M_1^{n-1}$ for $n>2$. In particular, we will focus on $R_3$ as it is the lowest order which can act as a coherence certifier.

\begin{table}[t]
\small
\begin{center}
\begin{tabular}{| c | c | c |}
\hline
$k$-coherence & $R_3$ Threshold & $R_3$ Best Known\\
\hline
$1$ & $1$ & $1$\\
$2$ & $5/4$ & $1.25$\\
$3$ & $179/96 \approx 1.86$ & $1.77$\\
\hline
\end{tabular}
\end{center}
\caption{The maximum values that $R_3(\rho, \ket{\chi})$ can attain, under any Hamiltonian, for any $\ket{\chi}$ and for any $\rho \in C_k$ as a function of $k$. As such, exceeding these values means that the state must be at least $(k+1)$-coherent. The middle column is an upper bound to this highest value obtained analytically. The last column is the highest value we found after conducting a thorough numerical optimisation.}
\label{tab:R3Bounds}
\end{table}

A central property of these functions is their convexity under the mixing of states
\be
R_n\left(\lambda\rho_1 + (1-\lambda)\rho_2\right) \le \lambda R_n\left(\rho_1\right) + (1-\lambda)R_n\left(\rho_2\right),
\label{eq:Convexity}
\ee
with the same $\ket{\chi}$ throughout, as proven in \appref{sec:ProofConvex}. As $C_k$ is itself convex, it is highly desirable for our certifier to also have this property as it implies that $R_n$ is maximised for pure states, {\it i.e.}
\be
\max_{\ket{\psi}\bra{\psi} \in C_k,\,\ket{\chi}} R_n(\ket{\psi}, \ket{\chi}) \ge \max_{\rho \in C_k,\,\ket{\chi}} R_n(\rho, \ket{\chi}),
\ee
where the ket in the first argument of $R_n$ stands for the corresponding pure state. Because of this, the maximum found for pure states also applies to mixed states directly.

Another useful feature of $R_n$ is that its maximum is reached when the measurement projector and the initial state are the same, {\it i.e.}
\be
\max_{\ket{\psi}\bra{\psi} \in C_k} R_n(\ket{\psi}, \ket{\psi}) \ge  \max_{\ket{\psi}\bra{\psi} \in C_k,\,\ket{\chi}} R_n(\ket{\psi}, \ket{\chi}).
\ee
This is not necessary for a coherence certifier, but is nevertheless desirable for two reasons. Firstly it aligns with the intuition of a Ramsey-like interferometer, where the highest contrast is obtained by projecting onto the initial state, which is also what was found in prior work where $\ket{\chi}$ was assumed to be the equal superposition state $\ket{W}$ \cite{VonPrillwitz2015}. Secondly it further simplifies calculating the threshold values, rather than maximising over the $4d$ real variables that define $\ket{\psi}$ and $\ket{\chi}$: it is enough to consider only the $d$ variables, $\psi_i\chi_i^*$, which can always be chosen such that they are real. This is proven in \appref{sec:ProofEqualStateMeas}.

Of particular importance is the need for $R_n$ to be hierarchical, such that it obeys the strict inequality
\be
\max_{\rho \in C_{k+1},\,\ket{\chi}} R_n(\rho, \ket{\chi}) > \max_{\rho \in C_k,\,\ket{\chi}} R_n(\rho, \ket{\chi}),
\ee
where the maximum for a given $k$ is known as the threshold value for $k+1$.
As proven in \appref{sec:ProofThresholdValues}, this holds for $k=1,2$ and $3$ independently of the dimension of the system Hilbert space. Observing a higher value than those thresholds, given in \tableref{tab:R3Bounds}, therefore proves that the state is at least $2, 3$, or $4$-coherent respectively.

The assumption so far is that the measurement is projective.
In practice, however, the realization of the unitary $U_r$ can be affected by noise,
and repetitions of the experiment that are required to obtain good statistics will suffer from fluctuations in $U_r$.

The signal on the measurement device will thus not reliably indicate projection onto the state $\ket{\chi}$, but rather randomly a projection onto one out of several states $\ket{\chi_j}$ occurring with probability $q_j$.  In this case the recorded interference pattern reads
\begin{align}
p(t)&=\sum_j q_j p_j(t),\quad\text{where}\\
p_j(t)&=\bra{\chi_j} \, U(t)\rho\, U^\dagger(t) \ket{\chi_j}\ ,
\end{align}
and the definition of moments given above in \eqref{eq:defineMoment} generalizes to
\begin{align}
M_n(\rho, \sigma_{\chi})&= \frac{1}{2\pi} \int_0^{2\pi} (\mbox{Tr}\ {e^{-i H t} \,\rho\, e^{i H t}\,\sigma_\chi})^n \,dt\, \nonumber
\end{align}
with $\sigma_\chi=\sum_jq_j\prj{\chi_j}$.
In exactly the same way that $R_n(\rho, \sigma_\chi)=M_n(\rho, \sigma_{\chi})/M_1(\rho, \sigma_{\chi})^{n-1}$ is convex in the first argument $\rho$ for any given $\sigma_\chi$, it is also convex in the second argument for any given $\rho$ such that
\be
R_n(\rho, \sigma_\chi) \le \sum_jq_jR_n(\rho,\ket{\chi_j}) \ ,
\label{eq:notprojective}
\ee
for any state $\sigma_\chi$ and convex decomposition into pure states $\sum_jq_j\prj{\chi_j}=\sigma_\chi$.
Since no projective measurement can overestimate the degree of coherence, no fluctuations in the realisation of such a measurement can result in a false positive either.

\section{Numerical threshold values}
\label{sec:num_threshold}

While the previous section details analytically proved results about the threshold values for $k$ up to $4$, we can go to much higher coherence levels numerically. We do this by maximising the value of $R_n$ over all $\rho \in C_k$ and all $\ket{\chi}$, for given values of $n$ and $k$. This problem is substantially simplified using the results of the previous section, which lets us set $\rho = \ket{\psi}\bra{\psi}$ and $\ket{\psi} = \ket{\chi}$ which only contain real coefficients in the eigenbasis of the Hamiltonian. We are confident that the results found this way are an excellent approximation of the true maxima as they are stable under different parameterisations of the problem and for different initial conditions in the numerical optimisation. These numerical results can also be compared to the upper bounds given by the analytic results, thereby illustrating how tight they are.

\begin{table}[h]
\footnotesize
	\begin{center}
    \begin{tabular}{|c|c|c|c|l|}
	\hline
    $R_n$ & $k$ & $R_n(\ket{\Psi_k})$ & $R_n(\ket{W_k})$ & $\boldsymbol{\Psi}_k$ \\
    \hline
    ~&~&~&~&~ \\[-1.2ex]
    $R_3$ & 2   & 1.25           & $1.25$ & $(0.50, 0.50)$ \\[1ex]
    ~     & 3   & 1.77           & $1.74$ & $(0.31, 0.38, 0.31)$ \\[1ex]
    ~     & 4   & 2.32           & $2.27$ & $(0.22, 0.28, 0.28, 0.22)$ \\[1ex]
    ~     & 5   & 2.88           & $2.80$ & $(0.17, 0.21, 0.23, 0.21, 0.17)$ \\[1ex]
    \hline
    ~&~&~&~&~ \\[-1.2ex]
    $R_4$ & 2   & 2.19           & 2.19 & $(0.50, 0.50)$ \\[1ex]
    ~     & 3   & 4.61           & 4.56 & $(0.32, 0.36, 0.32)$ \\[1ex]
    ~     & 4   & 8.02           & 7.90 & $(0.23, 0.27, 0.27, 0.23)$ \\[1ex]
    ~     & 5   & 12.42          & 12.21 & $(0.18, 0.21, 0.22, 0.21, 0.18)$ \\[1ex]
    \hline
    ~&~&~&~&~ \\[-1.2ex]
    $R_5$ & 2   & 3.94           & 3.94 & $(0.50, 0.50)$ \\[1ex]
    ~     & 3   & 12.39          & 12.28 & $(0.32, 0.36, 0.32)$ \\[1ex]
    ~     & 4   & 28.71          & 28.39 & $(0.24, 0.26, 0.26, 0.24)$ \\[1ex]
    ~     & 5   & 55.52          & 54.84 & $(0.19, 0.21, 0.21, 0.21, 0.19)$ \\
    \hline
    \end{tabular}
    \end{center}
	\caption{Numerical results for the first three hierarchical ratios for up to $5$-coherent states; showing their behaviour as coherence certifiers. The values of $R_n$ are given for the equal superposition state $\ket{W_k}$, and for the state $\ket{\Psi_k}$ which maximises the value (in all cases the basis for the projector $\ket{\chi}$ is equal to the state itself as we know that this maximises $R_n$). The states $\ket{W_k}$ and $\ket{\Psi_k}$ only have adjacent energy levels populated, spacing these levels out always results in a decrease in $R_n$ (unless they are spaced out equally in which case they are effectively adjacent levels for a different harmonic Hamiltonian). $\ket{\Psi_k}$ is found through numerical optimisation and is stable through different parametrisations of the problem and from different initial points. The amplitudes squared of $\ket{\Psi_k}$ are also listed as a vector to show how it differs from the uniform case of $\tfrac{1}{k}$.}
\label{tab:ratios}
\end{table}
	
These numerical results are listed in \tableref{tab:ratios}, which also shows the state $\ket{\Psi_k}$ that gives the maximum value of $R_n$ over all states in $C_k$, and how this value compares to the value given by the equally balanced state $\ket{W_k} = \tfrac{1}{\sqrt{k}}\sum_i^k \ket{i}$. These states are, surprisingly, not the same, although they both share the property of having $k$ adjacent basis states populated while the others have zero amplitude. $\ket{\Psi_k}$ has a concentration of population towards the middle of the occupied energy levels. One way to understand this is to note that interferences between basis states with small energy differences contribute more to $R_n$ than those with large energy differences. As the basis states in the middle of the spectrum are closer to more of the basis states, the function is maximised by populating them more than the others. This intuition is more visible in the re-parametrisation of $R_n$ done in \appref{sec:ProofThresholdValues}. Furthermore, the larger $k$ is and the smaller $n$ is, the more pronounced the difference between $\ket{W_k}$ and $\ket{\Psi_k}$ is.

\begin{figure}[h]
\centering
\includegraphics[width=0.95\columnwidth]{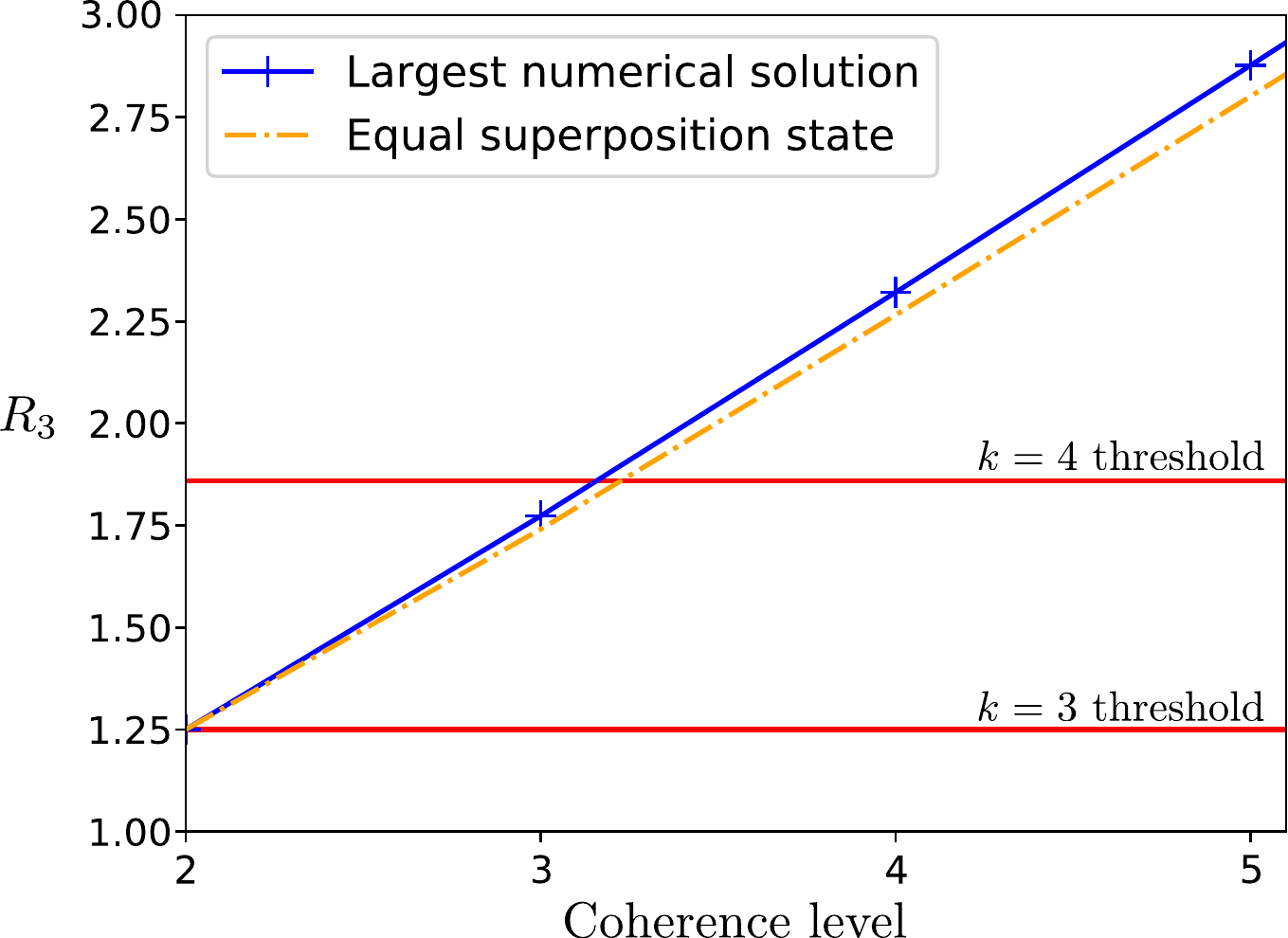}
\caption{Comparison of numerical and analytical threshold values. The crosses show the maximum value that we found for $R_3$ for $\rho\in C_k$ as a function of $k$. The solid blue line is a linear fit for these, showing how they are equally spaced. The dashed orange line shows what value the equal superposition state $\ket{W_k}$ has for the optimal measurement for comparison and is given by $\frac{4 + 5k^2 + 11k^4}{20k^3}$ (derived in \appref{sec:StraightLineR3}), which is asymptotically linear. The horizontal lines are the analytic threshold values. For the 2-coherent case, the equal superposition and optimal states overlap, and lie immediately below the threshold for certifying 3-coherence. For the 3-coherent case and higher, there is a finite but small gap between the equally balanced and optimal states. The threshold for 4-coherence also does not lie exactly above the maximum for 3-coherence, but the gap is again very small and, as we are lower bounding the amount of coherence present, this only means that $R_3$ is occasionally too cautious.}
\label{fig:BoundsComparison}
\end{figure}

In all cases of interest, however, the difference in the $R_n$ value between $\ket{\Psi_k}$ and $\ket{W_k}$ is relatively small, which can be seen in \figref{fig:BoundsComparison}. This figure also compares these to the analytic thresholds which shows how tight they are. Furthermore, the maximal values grow linearly (tested up to $k=30$, not shown on the graph). This constant interval means that $R_3$ would also be able to distinguish between more highly coherent states. The functions $R_4$ and $R_5$ seem to have even faster growth, potentially making them more useful in such circumstances, although the additional experimental difficulty in accurately reconstructing higher moments should not be neglected \cite{Flusser2009}.

%%%%%%%%%%%%%%%%%%%%%%%%%%%%%%
\section{Verification of $k$-coherence in the presence of imperfections}
\label{sec:exp}

In this section we demonstrate that the present approach can verify coherence properties, even in the presence of substantial imperfections in the projective measurement and that coherence can be detected even in highly mixed states.

\subsection{Measurement tolerance}
\label{sec:tolerance}

Having proved that an imperfect measurement will never overestimate the coherence of a state, it is important to demonstrate that it does not underestimate it too strongly either. Therefore, we quantify this implication of measurement imperfections here. To achieve this, we produce a sample of random faulty measurements and estimate the deviation from perfect measurement required to reduce the value of the maximum $k$-coherent state below the threshold below which $k$-coherence is not verified anymore.

We define the states $\ket{\chi_k(\tau)}$ that define a projective measurement in terms of a random Hamiltonian $H_r$ via the relation
\begin{equation}
	\label{eq:meas_evolution}
	\ket{\chi_k(\tau)} = \mathcal{U}(\tau) \ket{\Psi_k} \coloneqq e^{iH_r\tau}\ket{\Psi_k}\ ,
\end{equation}
with $\ket{\Psi_k}$ given in Tab.~\ref{tab:ratios};
the random Hamiltonians $H_r$ are drawn from the Gaussian Unitary Ensemble (GUE)~\cite{ref:Fyodorov}.

The degree to which the projective measurement deviates from the ideal measurement can be quantified by the norm
\begin{equation}
	\label{eq:meas_norm}
	D(\tau) \coloneqq ||\ket{\chi_k(\tau)} - \ket{\chi_k(0)}|| \equiv \sum\limits_{i=1}^d \left[ \mathcal{U}_{ij}(\tau) \chi_j - \chi_i \right]^2\ ,
	\end{equation}
for each realisation of $H_r$.

\begin{figure}
\centering
    \includegraphics[scale=0.5]{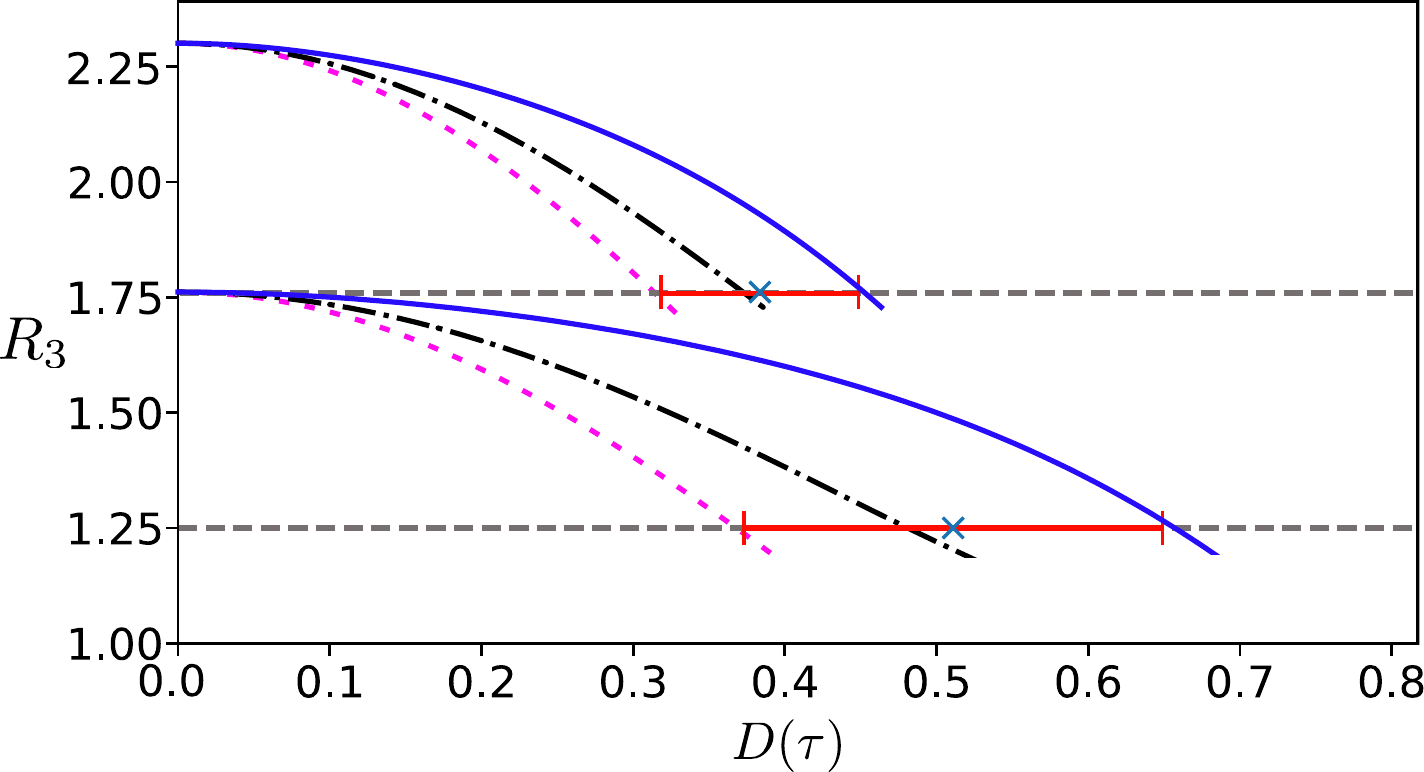}
\caption{Ensemble average (black) of $R_3$ for the states $\ket{\Psi_k}$ ($k=4$ at the top, $k=3$ at the bottom) obtained with faulty measurements as function of the measurement deviation $D$ defined in~\eqref{eq:meas_norm}.
The standard deviation of the random distribution is depicted with solid (blue) and dotted (pink) lines centred around the average. The values of $D$, for which the threshold values (Tab.~\ref{tab:R3Bounds}) are reached, are depicted in turqoise crosses (ensemble average) with solid horizontal red lines for the width of the distribution.}
\label{fig:tolerance}
\end{figure}

\figref{fig:tolerance} depicts the ensemble average of $R_3(\ket{\Psi_k},\ket{\chi_k(\tau)})$ with the average performed over $100$ random Hamiltonians as function of $D(\tau)$ with black lines for $k=3$ and $k=4$.
The blue and pink lines depict the width of the underlying distribution, and the horizontal black dashed lines depict the threshold values for the detection of 3-coherence and 4-coherence.
As one can see, a substantial value of $\tau$ is required before the recorded values of $R_3$ drop below the threshold values.
As one might have expected the verification of 3-coherence can tolerate a large amount of deviations, but even for  the verification of 4-coherence, a deviation $D\le 0.3$ is typically good enough.

\subsection{Decoherence tolerance}
\label{sec:decoh}

Since our central aim is the ability to verify coherence in the presence of experimental imperfections, the big remaining question is on the degree of decoherence that can be present, before the present criteria fail to verify a desired level of coherence. Repetitions of the experiment may result in some instances of initially mixed states which are then evolved through the system. The decoherence effects of such a faulty state preparation is to reduce the visibility of the interference pattern, thus rendering the task of bounding coherence more challenging.

We explore the impact of decoherence by introducing the Werner-like state~\cite{ref:Werner}
\begin{equation}
	\label{eq:werner}
	\rho_W = (1-\lambda)\ket{W_k}\bra{W_k} + \frac{\lambda}{k} \mathbb{I}_k\ ,
\end{equation}
and exploring the ability of the ratios to distinguish its level of coherence. The Werner-like state is 
given by a mixture of the equal superposition $k$-coherent state and the totally incoherent state $\mathbb{I}_k/d$.
The degree of mixedness is varied with the parameter $\lambda \in [0,1]$. For $\lambda = 0$, the system is pure and $k$-coherent, while $\lambda = 1$ corresponds to a completely mixed state.
Therefore, there must be a theoretical upper bound $\lambda_{\text{dec}}(q)$ above which the system is in $C_q$, but not $C_{q+1}$, and as the noise increases further there must be another bound above which the coherence drops further. For a $k$-dimensional system, these bounds $\lambda_{\text{dec}}(q)$ are given by
\begin{equation}
	\label{eq:decoh_thr}
	\lambda_{\text{dec}}(q) = \frac{k-q}{k-1}, \quad 1 \leq q \leq k,
\end{equation}
as proved in \appref{sec:decoh_thr} and also discussed in Ref.~\cite{Ringbauer2018}.
Similarly, we can define threshold values $\lambda_{\text{thr}}^{(n)}(q)$ at which a given certifier $R_n$ fails to verify $(q+1)$-coherence in a system from its interference pattern.
The values $\lambda_{\text{thr}}^{(n)}(k-1)$ at which a given certifier $R_n$ fails to identify $k$-coherence are depicted in \tableref{tab:r3_dthr} for $R_3$, $R_4$ and $R_5$, and numerical expressions for $\lambda_{\text{dec}}(q)$ are given for comparison.

As one can see, the threshold values for the detection of $k$-coherence are larger the smaller $k$ is.
$k$-coherence can thus be identified for rather strongly mixed states as long as $k$ is sufficiently low.
$R_5$ can identify coherence for larger values of $\lambda$ ({\it i.e.} more strongly mixed states) than $R_4$ for any value of $k$, and $R_4$ outperforms $R_3$ in the same sense.
If a given $R_n$ fails to verify $k$-coherence in a strongly mixed state, one can thus resort to a certifier $R_n$ with a larger value of $n$, and find better performance.
Even for $R_5$, however, the threshold value $\lambda_{\text{dec}}(k-1)$ is about $50\%$ larger than $\lambda_{\text{thr}}^{(5)}(q)$, and higher moments would be required in order to identify the $k$-coherence in very strongly mixed states. 

\begin{table}
\small
	\begin{center}
   \begin{tabular}{|c|cccccccc|}
	\hline
    $k$ & $3$ & $4$ & $5$ & $6$ & $7$ & $8$ & $9$ & $10$ \\
    \hline
    $\lambda_{\text{thr}}^{(3)}(k-1)$ & $0.18$ & $0.13$ & $0.10$ & $0.08$ & $0.06$ & $0.06$ & $0.05$ & $0.04$ \\
    $\lambda_{\text{thr}}^{(4)}(k-1)$ & $0.28$ & $0.19$ & $0.14$ & $0.11$ & $0.09$ & $0.08$ & $0.07$ & $0.06$\\
    $\lambda_{\text{thr}}^{(5)}(k-1)$ & $0.33$ & $0.22$ & $0.16$ & $0.13$ & $0.11$ & $0.09$ & $0.08$ & $0.07$\\
    $\lambda_{\text{dec}}(k-1)$ & $0.5$ & $0.33$ & $0.25$ & $0.2$ & $0.17$ & $0.14$ & $0.13$  & $0.11$\\
    \hline
    \end{tabular}
    \end{center}
\caption{Numerical expressions for decoherence thresholds for the state $\rho_W$ from \eqref{eq:werner}, for $R_3, R_4$ and $R_5$ between consecutive levels of coherence for $k = 3$ to $10$.}
\label{tab:r3_dthr}
\end{table}

\subsection{Best approximations of interference pattern}

In addition to the thresholds $\lambda_{\text{dec}}(q)$ and $\lambda_{\text{thr}}^{(n)}(k-1)$ discussed above, there is also the threshold $\lambda_{\text{patt}}(q)$ at which a given interference pattern no longer allows to verify $q$-coherence.  As $R_n$ is a scalar functional of the interference pattern, it can contain up to as much information as the pattern itself, and a small difference between $\lambda_{\text{patt}}$ and $\lambda_{\text{thr}}^{(n)}$ indicates that only little information is lost by looking at the ratio of specific moments instead of the full interference pattern.

These thresholds, for all $n$, satisfy the relation
\begin{equation}
	\label{eq:decoh_inequality}
	0 < \lambda_{\text{thr}}^{(n)}(q) \leq \lambda_{\text{patt}}(q) \leq \lambda_{\text{dec}}(q) < 1,
\end{equation}
for all $n$. The value of $\lambda_{\text{patt}}(q)$ is strongly dependent on the measurement projection $\ket{\chi}\bra{\chi}$, and we show in \appref{sec:decoh_thr} that the threshold values $\lambda_{\text{patt}}(q)$ and $\lambda_{\text{dec}}(q)$ nevertheless coincide for Werner-like states, with a projection onto the equal superposition $k$-coherent state $\ket{W_k}$.
Strikingly, this verifies that in this case a single interference pattern can provide enough information for a complete classification of $q$-coherence.

For $\lambda \geq \lambda_{\text{patt}}(q)$, a $q$-coherent state is mixed enough to produce a pattern $p(t)$ which can be reproduced by states of lesser coherence. 
Patterns $p(t)$ resulting from states with $\lambda < \lambda_{\text{patt}}(q)$, on the other hand, cannot be reproduced by states in $C_{q-1}$. In order to exemplify the differences in interference patterns that the present criteria aim at identifying, for a given interference pattern $p(t)$ produced by a $k$-coherent state with $k > q$, we introduce the best $q$-approximation $\bar{p}_{q}(t)$ to $p(t)$, as the interference pattern resulting from $q$-coherent states only with minimal deviation from $p(t)$.

\begin{figure}[t]
\centering
    \includegraphics[scale=0.5]{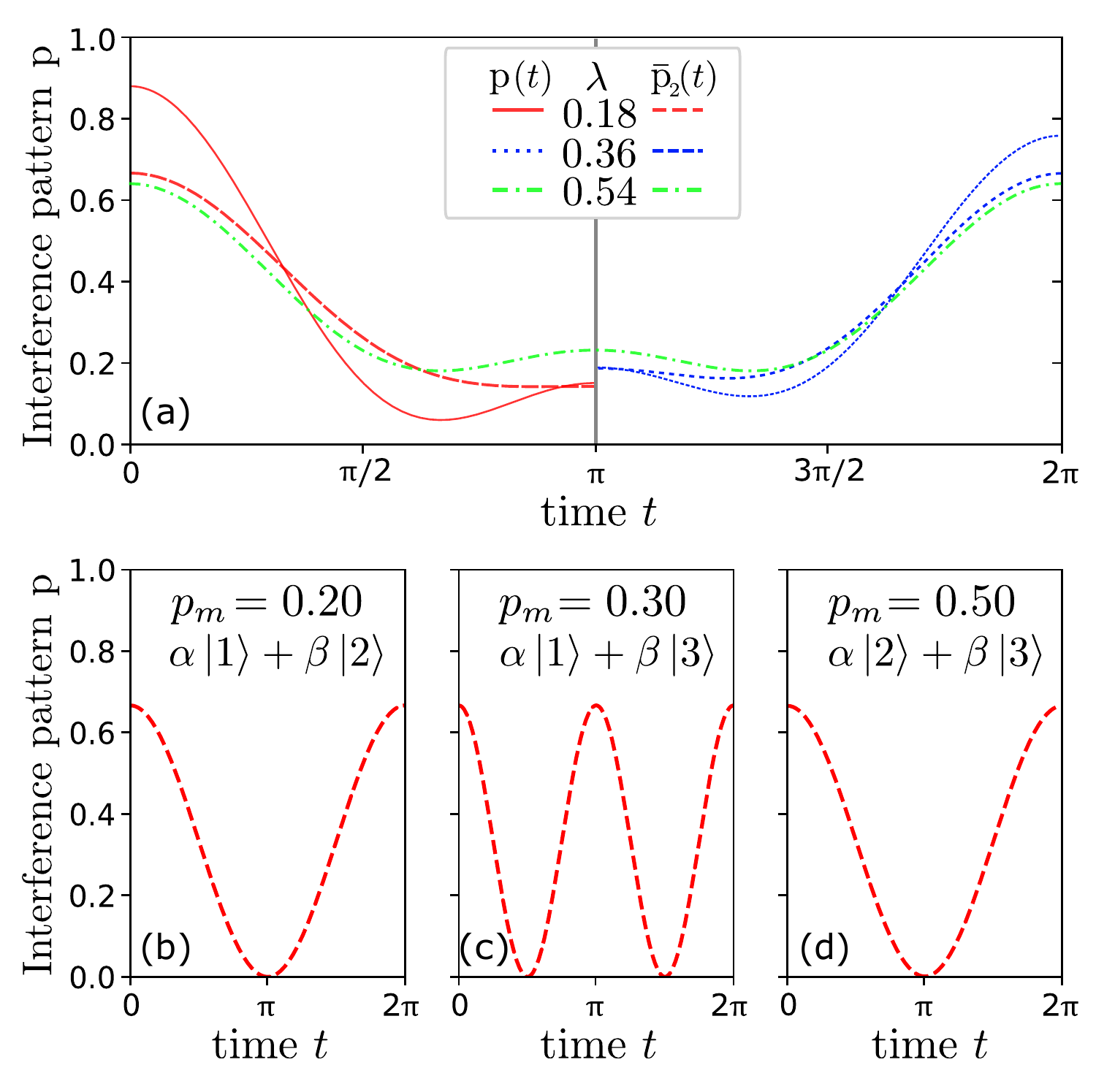}
\caption{(a) Interference patterns $p(t)$ of $\rho_W \in C_3$ in a 3-dimensional space with different values of $\lambda$ projected under optimal measurement state $\ket{W_3}$ (solid/dotted curves), along with their best approximations $\bar{p}_{2}(t)$ (dashed curves) reproduced by states in $C_2$. The $R_3$ values of the states are, $1.26, 0.88, 0.60$, with increasing $\lambda$, so the system corresponding to $\lambda=0.18$ can be certified by $R_3$ as 3-coherent. \\
(b)-(d) Three linearly independent 2-coherent states that, when mixed with the given probabilities $p_m$, provide the best approximation of $\rho_W$ at $\lambda=0.18$.
}
\label{fig:pattdecomp}
\end{figure}

In \figref{fig:pattdecomp}, we focus on $R_3$ and we investigate the ability to detect 3-coherence on states $\rho_W \in C_3$ with optimal projection. The patterns corresponding to $\rho_W$ are plotted for $\lambda =  0.18, 0.36$ and $0.54$ along with their best approximations $\bar{p}_{2}(t)$. As long as $\lambda < \lambda_{\text{dec}}(3)=\frac{1}{2}$, which is the case for the red and blue curves (corresponding to $\lambda = 0.18, 0.36$ respectively), the pattern cannot be reproduced by states in $C_2$, as expected, since $\lambda_{\text{patt}}(3) = \lambda_{\text{dec}}(3)$. The green pattern ($\lambda = 0.54$) can be reproduced by states in $C_2$ exactly, because in this case $\lambda > \lambda_{\text{dec}}(3)$. If the projection was sufficiently far from the optimal, the red and blue patterns would also be exactly reproducible by patterns of 2-coherent states. The red pattern corresponds to $\rho_W(\lambda=0.18)$, and has a value of $R_3$ (when maximised over $\ket{\chi}$) of $1.26$, which lies above the threshold given in \tableref{tab:R3Bounds} to certify a state as $3$-coherent. The bottom part of the plot gives the patterns of the three 2-coherent states which, when mixed, provide the best approximation to the red $\rho_W$ pattern. Three is the minimum number of basis states required to form the 3-dimensional Werner-like state. For $\lambda>\frac{2}{3}$, the patterns could also be decomposed simply into incoherent states, as \eqref{eq:decoh_thr} indicates.

%%%%%%%%%%%%%%%%%%%%%%%%%%%%%%
\section{Conclusion}
\label{sec:conclusion}

Despite the numerous similarities between the theories of entanglement and coherence, the equality in operation required for creation and verification of quantum coherence defines a crucial difference between those two theories. Our proposed solution relies on easily observable quantities such that an imperfectly implemented verification protocol can never overestimate the degree of coherence. As such, it offers very practical and robust avenue to rigorously verify coherence properties beyond the two-level setting.

Beyond the fundamental question `{\it when is a triple-slit interference pattern so washed out, that one can not recognize it anymore?}', the ability to verify the number of states contributing to a coherent superposition has also very practical applications in the verification that a potential quantum device is actually able to operate in the quantum regime that it is supposed to.

\section{Acknowledgments}
We are grateful for stimulating discussions with Nicky Kai Hong Li that motivated this work. B.D. acknowledges funding from the Engineering and Physical Sciences Research Council (EPSRC UK) administered by Imperial College London via the Postdoctoral Prize Fellowship program for the core duration of this work; and funding from  Austrian Science Fund (FWF): P 30947 for the closing stages. N.K. acknowledges funding from the EPSRC UK through the Controlled Quantum Dynamics Centre for Doctoral Training for the closing stages.

\small
\bibliography{tidy_bib}

%%%%%%%%%%%%%%%%%%%%%%%%%%%%%
%%%%%%%%%%%%%%%%%%%%%%%%%%%%%
\onecolumngrid
\newpage
\appendix

%%%%%%%%%%%%%%%
\section{Proof that $R_n$ is convex in either argument}
\label{sec:ProofConvex}
To prove that $R_n$ is convex under the mixing of states it suffices to show that 
\be
R_n\left(\lambda\rho_1 + (1-\lambda)\rho_2, \ket{\chi}\right) \le \lambda R_n\left(\rho_1, \ket{\chi}\right) + (1-\lambda)R_n\left(\rho_2, \ket{\chi}\right),
\label{eq:ConvexityAppendix}
\ee
for all pairs of states $\rho_1, \rho_2$, for all projectors $\ket{\chi}\bra{\chi}$, and for all $\lambda\in[0, 1]$.

This property holds for the moments themselves, which are convex and positive by construction. Products and sums of such functions stay convex, but this is not necessarily the case for ratios of them. We prove that this particular function is indeed convex, for $n \ge 2$, by taking the second derivative of \eqref{eq:ConvexityAppendix} with respect to $\lambda$ and showing that it is always non-negative. 

This second derivative is
\begin{align}
&\partial_{\lambda}^2R_n = \frac{M_1^{3n-5}}{M_1^{4n-4}} \big[ M_1^2 \partial_{\lambda}^2 M_n - 2(n-1)M_1(\partial_{\lambda}M_1) \partial_{\lambda}M_n + n(n-1)(\partial_{\lambda} M_1)^2 M_n \big]. \nonumber
\label{eq:secondDeriv}
\end{align}
Denoting the integrand of $M_n$ in \eqref{eq:defineMoment} by $p^n$ and the time average by $\langle \cdot \rangle$, allows the derivatives to be calculated according to
\begin{align}
\partial_{\lambda} M_n &= \langle n  (\partial_{\lambda}p)  p^{n-1}\rangle, \nonumber \\ 
\partial_{\lambda}^2 M_n &= \langle n(n-1) (\partial_{\lambda} p)^2  p^{n-2} \rangle. \nonumber
\end{align}
Substituting these expressions into \eqref{eq:secondDeriv} gives
\begin{align}
&\partial_{\lambda}^2R_n =\frac{M_1^{3n-5}}{M_1^{4n-4}} \langle n(n-1)p^{n-2} \left[p \langle \partial_{\lambda}p \rangle - \langle p \rangle \partial_{\lambda}p \right]^2 \rangle,
\end{align}
where the fraction at the front is non-negative, as is the squared term in the time average and its pre-factor (for $n\ge2$), thereby showing that $R_n$ is convex as desired.

%%%%%%%%%%%%%%%
\section{Proof that $R_n$ is maximised for equal preparation and projection}
\label{sec:ProofEqualStateMeas}

We begin by noting that the expression for the probability distribution in \eqref{eq:defineMoment} for pure states is given by the double sum
\begin{align}
p(\ket{\psi}, \ket{\chi}, t) =  \sum_{p, q} \chi_p^* \psi_p \psi_q^* \chi_q e^{-i(p - q)t},
\end{align}
where $\ket{\psi} = \sum_p \psi_p \ket{p}$, $\ket{\chi} = \sum_q \phi_q \ket{q}$ and the basis states are eigenkets of the Hamiltonian of \eqref{eq:BasicHamiltonian} $H = \sum_n  n \ket{n}\bra{n}$. By defining $\psi_p \chi_p^* = \alpha_p e^{i \phi_p}$, $\phi_{pq} = \phi_p - \phi_q$ and $\omega_{pq} = p - q$ this can be recast as
\begin{align}
p(\ket{\psi}, \ket{\chi}, t) =  \sum_p \alpha_p^2 + 2\sum_{p>q}  \alpha_p \alpha_q \cos(\omega_{pq} t + \phi_{pq}),
\label{eq:CosProbability}
\end{align}
where the $\alpha$ are real and non-negative by construction.

We now show that the maximum of this over $k$-coherent $\ket{\psi}$ and any $\ket{\chi}$ is reached when the phases $\phi_{pq}$ are all zero, for all $k$. Firstly, because integrating cosines over an integer number of periods gives zero, the first moment is independent of them,
\begin{align}
M_1 = 2 \pi \sum_p \alpha_p^2.
\end{align}
It is therefore clear that changes in $\phi_{pq}$ (arising from different phases between the state and the projector) affect the numerator of $R_n$ but not the denominator. The terms of $M_{>1}$ which depend non-trivially on the phases are inside the integral over time and are of the form
\begin{align}
\int_0^{2\pi} \left(\sum_{p>q} \alpha_p\alpha_q \cos(\omega_{pq} t + \phi_{pq})\right)^m \,dt.
\end{align}
To see which terms do not vanish when integrated over, it is useful to look at the products of cosines individually
\begin{align}
\int_0^{2\pi} &\alpha_{p_1}\alpha_{q_1} \cos(\omega_{p_1q_1} t + \phi_{p_1q_1})\times\alpha_{p_2}\alpha_{q_2}  \cos(\omega_{p_2q_2} t + \phi_{p_2q_2})\times...\,dt.
\end{align}
which can themselves be expanded into a sum of cosines, where each term is of the form
\begin{align}
\propto \int_0^{2\pi} \cos\left[(\omega_{p_1q_1} \pm \omega_{p_2q_2}...)t + \phi_{p_1q_1} \pm \phi_{p_2q_2}...\right]\,dt.
\end{align}
If the sum (for the different permutations of signs) of frequencies do not sum to $0$, then the integral vanishes. If they do sum to $0$, the term is proportional to the cosine of the sum (for the different permutations of signs) of the phases. One of the solutions which maximises this is to pick all the $\phi_{pq}=0$, which simultaneously maximises every such integral no matter the number of terms or the sign configuration. This itself increases $M_n$ and therefore the value of $R_n$.\\

In this case that there are no relative phases, $R_n$ can be written in terms of a simplified \eqref{eq:CosProbability} as
\begin{align}
R_n(\ket{\psi},\ket{\chi}) = \frac{\int_0^{2\pi}\left(\sum_p \alpha_p^2 + 2 \sum_{p>q} \alpha_p\alpha_q \cos(\omega_{pq} t) \right)^n \,dt} {2\pi\left(\sum_p \alpha_q^2\right)^{n-1}}.
\label{eq:RnNoPhase}
\end{align}
From this it can be seen that the mapping $\alpha_p \to x\alpha_p$ changes the function $R_n\to x^2R_n$. It is therefore desirable to scale the $\alpha$ to be as large as possible. The extent to which this can be done is bounded by the normalisations of the states, using Cauchy-Schwarz we can express this as:
\begin{align}
\left(\sum_p \alpha_p\right)^2 = \left(\sum_p \psi_p \chi_p \right)^2  &\le \left(\sum_p \psi_p^2 \right) \; \left(\sum_p \chi_p^2 \right) = 1, \\
\implies \sum_p \alpha_p &\le 1, \nonumber
\end{align}
Furthermore, any set of $\{\alpha_p\}$ that satisfy this bound can be realised by the normalised states $\ket{\psi}$, $\ket{\chi}$ by picking their amplitudes according to $\psi_p = \chi_p = \sqrt{\alpha_p}$.

Taking a step back, what we have shown by parametrising the function $R_n$ in terms of $\{\alpha_p, \phi_p\}$, is that the maximum of $R_n$ occurs when $\phi_p = 0$ and $\sum \alpha_p$ = 1. These two conditions are equivalent, in terms of the physical state and measurement projector, to having $\ket{\psi}= \ket{\chi}$. Thus, we know that $R_n$ is maximised when the input state is pure and the projective measurement is equal to it, thereby greatly shrinking the space over which we have to optimise. Note that this is not the same as the subtly different question of whether the optimal $\ket{\chi}$ that should be picked for a given $\ket{\psi}$ is for them to be the same. Here, we are only interested in the overall bound $R_n$ can have over any input state with a fixed $k$-coherence.

%Therefore, in the case that $\sum\alpha_p = A \le 1$, the value of $R_n$ can be increased by instead choosing $\alpha_p^\prime = \alpha_p A^{-1}$. Furthermore this bound is tight in that there always exist a normalised state $\ket{\psi}$ and measurement basis $\ket{\chi}$ which leads to $(\sum \alpha_p A^{-1})^2 = 1$ for any given set $\{\alpha_p\}$, namely, $\psi_k = \chi_k = \sqrt{\alpha_p A^{-1}}$. Indeed, from Cauchy-Schwarz, the only choice of $\ket{\psi}$ and $\ket{\chi}$ which lead to $\sum\alpha_p = 1$ is when $\ket{\psi} = \ket{\chi}$.
% This condition is equivalent to saying that, at the maximum value of $R_n$ for a fixed $k$-coherence, the input state is pure and identical to the projective measurement. Mathematically this is the requirement that in \eqref{eq:RnNoPhase} the $\alpha_p$ are all non-negative and sum to $1$. Note that this is a slightly different statement to saying that for every pure state $R_n$ is maximised by picking the projector to be the same as the state.

%%%%%%%%%%%%%%%
\section{Derivation of analytic threshold values}
\label{sec:ProofThresholdValues}
We now compute the maximum of \eqref{eq:RnNoPhase} for $n=3$ as a function for $k$ where $\ket{\psi}\bra{\psi} \in C_k$. For $k=2$, this is easily done by using the previously found constraint of $\sum \alpha_p = 1$. We denote the two non-zero $\alpha$'s as $x$ and $1-x$ and can perform the integration over time explicitely to arrive at
\begin{equation}
\max_{\rho \in C_2, \ket{\chi}} R_3(\rho, \ket{\chi}) = \max_{x\in[0,1]} \frac{1 + 2x(x-1)(5x^2-5x+2)}{1 + 2x(x-1)}.
\end{equation}
The right hand side is easily solved analytically and gives a value of $5/4$. Therefore, measuring an $R_3$ of greater than that value implies that the state must be at least $3$-coherent.\\

To deal with higher $k$, it is highly advantageous to reparametris the optimsation problem. the starting point is \eqref{eq:RnNoPhase} and we now make another simplification in the notation by grouping together terms with the same frequency $\omega_{pq} = p - q$. This allows the sum over the cosines to be expressed as
\begin{align}
\sum_{p>q} \alpha_p\alpha_q \cos(\omega_{pq} t) &= \sum_n D_n \cos(\omega_n t)
\end{align}
where the new variables are given by
\begin{align}
\quad D_n &= \sum_p \alpha_{p+n} \alpha_p, \quad \omega_n = n,
\label{eq:DConstraints}
\end{align} 
which also lets us rewrite the term $\sum_p \alpha_p^2 = D_0$, thereby unifying the notation. We also recall that, from previous arguments, that $\sum_p \alpha_p = 1$ for the maximum of the function. Using this notation in \eqref{eq:RnNoPhase} for the case $n=3$ we obtain
\begin{align}
R_3(\ket{\psi},\ket{\chi}) = \int_0^{2\pi} \big[&D_0 + 6 \sum_i D_i \cos(\omega_i t) + \frac{12}{D_0} \sum_{ij} D_i D_j \cos(\omega_i t) \cos(\omega_j t) +\\
&\frac{8}{D_0^2}\sum_{ijk} D_i D_j D_k \cos(\omega_i t) \cos(\omega_j t) \cos(\omega_j t)\big]\;dt\nonumber
\end{align}
Performing the integrals in the way described earlier, only terms where the $\omega$ sum to $0$ contribute, which yields
\begin{align}
R_3(\ket{\psi},\ket{\chi}) &= 6 D_0 \left(\frac{1}{6} + \sum_{i j} \frac{D_i D_j}{D_0^2}\delta_{i j} + \sum_{i j k} \frac{D_i D_j D_k}{D_0^3}\sigma_{i j k} \right) \nonumber\\
&= D_0 \left(\frac{1}{6} + \sum_i \DDi^2 + \sum_{i j k} \DDi \DDj \DDk \sigma_{i j k} \right)
\label{eq:R3InDs}
\end{align}
where $\sigma_{ijk}$ is a phase matching condition which is $1$ if $i + j = k$ and $0$ otherwise, and $\DDi = D_i/D_0$.

Finding the maximum value of $R_3$ over all $k$-coherent states as a function of $k$ has proved very difficult. What we have found is a method to calculate an analytic upper bound for this quantity for a given $k$. We have evaluated this bound for small $k$ and, although the method is applicable in general, it may be too laborious to be practical for high $k$.

The key idea is to treat the $\{\DDi\}$ as independent variables to optimise over and $D_0$ as a `free' parameter. \eqref{eq:DConstraints} is used to form linear constraints on the $\{\DDi\}$, which forms an outer approximation to the physically allowed region for a choice of $D_0$. We then show that in this region $R_3$ has a positive definite Hessian, which implies that for any line cutting through this region, the maxima of the function must be reached where the line crosses the bounding surface. Therefore, the maximum value is attained at one of the vertices. As this region is defined by linear constraints it is a polytope, and hence has only a finite number of vertices which can be individually evaluated to see which produces the largest value of $R_3$. The remaining step is then optimise over $D_0$, which is easily done numerically as the problem is reduced to finding the turning points of a quotient of low order polynomials in one variable.

Although we could not show that the Hessian is positive in general, we do find that it is in all the cases of interest. For convenience, it is useful to list its components here. These are the derivatives of $R_3$, which are given by
\begin{align}
\partial_{\tilde{D}_a} R_3 &= 6 D_0 \left(2D_a +  2 \sum_{j k} \DDj \DDk \sigma_{a j k}+ \sum_{i j} \DDi \DDj \sigma_{i j a}\right) \nonumber\\
\partial_{\tilde{D}_a} \partial_{\tilde{D}_a} R_3 &= 12 D_0 (1 + D_{2a}) \\
\partial_{\tilde{D}_b} \partial_{\tilde{D}_a} R_3 &= 12 D_0 (D_{a + b} + D_{|a-b|}).
\end{align}

We now apply the method outlined above to $k=3$, treating the case where the dimension of the Hamiltonian is truncated at $d=3$ and where it is unbounded separaly. We also calculate the case $k=4, d=4$ to show that the method can be applied to higher coherence levels.

%%%%%%%%%%
\subsection*{$k=d=3$}
For states which are at most $3$-coherent in a $3$-dimensional Hamiltonian, the variables are explicitely given by:
\begin{align*}
D_0 &= \alpha_1^2 + \alpha_2^2 + \alpha_3^2\\
D_1 &= \alpha_1 \alpha_2 + \alpha_2 \alpha_3\\
D_2 &= \alpha_1 \alpha_3 \\
1 &= \alpha_1 + \alpha_2 + \alpha_3.
\end{align*}
From this we can write some inequalities which constrain the allowed values. Firstly, as the $\alpha$'s are all positive we have that $0\le D_0$ and $0\le\DDi$. Secondly, the triangle inequality implies that
\begin{equation}
\frac{1}{d} \le  D_0 \le 1.
\label{d0bounds}
\end{equation}
The first non-trivial constraint comes about from the same starting point
\begin{align}
1 &= (\alpha_1 + \alpha_2 + \alpha_3)^2 \nonumber \\
&=  D_0 + 2 D_1 + 2 D_2 \nonumber \\
&=  D_0 \left(1 + 2 \sum_i \DDi \right).
\label{d3Equality}
\end{align}
From these two relations we upper bound the maximum values of any $\DDi$
\begin{align}
\frac{1}{d} \left(1 + 2 \sum_i \DDi \right) \le 1 \nonumber \\
\sum_i \DDi \le \frac{d - 1}{2} \nonumber \\
\DDi \le \frac{d - 1}{2}.
\label{Dibound}
\end{align}
Other inequalities can be obtained by considering well chosen sums of squares, the three useful ones are listed here. Firstly
\begin{align}
(\alpha_1 - \alpha_3)^2 + \alpha_2^2 \ge 0 \nonumber \\
 D_0 - 2 D_2 \ge 0 \nonumber \\
1 - 2 \tilde{D}_2 \ge 0.
\label{d3Ineq1}
\end{align}
Changing the sign gives a different inequality
\begin{align}
(\alpha_1 + \alpha_3)^2 + \alpha_2^2 \ge \tfrac{1}{2} \nonumber \\
 D_0 + 2 D_2 \ge \tfrac{1}{2} \nonumber \\
 D_0(1 + 2 \tilde{D}_2) \ge \tfrac{1}{2},
\label{d3Ineq2}
\end{align}
where the triangle inequality is used in the first line. Lastly, there is
\begin{align}
(\alpha_1 - \alpha_2 + \alpha_3)^2 \ge 0 \nonumber \\
 D_0 - 2 D_1 + 2 D_2 \ge 0 \nonumber \\
1 - 2 \tilde{D}_1 + 2 \tilde{D}_2 \ge 0.
\label{d3Ineq3}
\end{align}
The last three equations (for fixed $ D_0$) define a triangular region of interest, while \eqref{d3Equality} is a line that cuts through it. They can be expressed as succinctly as
\begin{align}
\max\left(\frac{1 - 2 D_0}{4 D_0}, 0\right) &\le \tilde{D}_2 \le \frac{1}{2}\nonumber \\
\label{d3complete}
0 &\le \tilde{D}_1 = \frac{1 -  D_0}{2  D_0} - \tilde{D}_2 \le 1\\
0 &\le 1 - 2 \tilde{D}_1 + 2 \tilde{D}_2 \nonumber
\end{align}

In order to be sure that the maxima of the function in this region is located at the vertices, we need the Hessian, which is
\[ \left(\begin{array}{c c}
1 + D_2 & D_1\\
D_1 & 1
\end{array}\right), \]
which is strictly positive definite everywhere in the allowed region. Therefore, the only points that need to be examined are the vertices of the polytope (in this case, just a line) defined by \eqref{d3complete} for the valid range of $ D_0$. It therefore just remains to find these vertices by solving these equations on the boundary in the $\tilde{D}_1 - \tilde{D}_2$ plane, which depends on the value of $ D_0$. They can be summarised as
\[ \begin{array}{c c c | c}
\tilde{D}_1 & \tilde{D}_2 &  D_0 & \max R_3 \\
\hline
\frac{1 -  D_0}{2  D_0} & 0 & \frac{1}{2}\le D_0\le1 & 1.25\\
\hline
\frac{1 - 2  D_0}{2  D_0} & \frac{1}{2} & \frac{1}{3} \le  D_0 \le \frac{1}{2} & 1.58 \\
\frac{1}{4  D_0} & \frac{1 - 2  D_0}{4  D_0} & \frac{1}{3} \le  D_0 \le \frac{1}{2} & 1.86
\end{array} \]
where the largest values of $R_3$ over all $ D_0$ in the allowed range are also given. From this we can conclude that if $R_3$ is larger than $1.86$ we can certify that the state is not a 3-coherent state lying in adjacent energy levels of an SHO. For comparison, the perfectly balanced state gives $1.74$ and the largest value we could fine numerically was $1.77$. The largest value found for a 4-coherent state (that we want to distinguish from) is $2.32$, while for a 2-coherent state it is $1.25$.

%%%%%%%%%%
\subsection*{$k=3, d\ge3$}
We now remove the restriction on the dimension and instead restrict ourselves to a 3-level state, which is to say that only 3 of the $\alpha$'s are non-zero. Without loss of generality, we have as the three populated levels $1, p, q$ with $1 < p < q$. This means that the only non-zero variables are $\alpha_1, \alpha_p, \alpha_q$, which gives
\begin{align}
D_0 &= \alpha_1^2 + \alpha_p^2 + \alpha_q^2, \\
D_{p-1} &= \alpha_1 \alpha_p, \\
D_{q-p} &= \alpha_p \alpha_q, \\
D_{q-1} &= \alpha_1 \alpha_q
\end{align}
with the assumption that $p-1 \neq q-p$. If these are equal, then the energy levels are equally spaced and we are back to the 3-level case considered in the first instance. As before, we now find inequalities on the $\tilde{D}$'s to define a volume. As each one only contains a single term, this can be done for each independently by considering
\begin{align}
(\alpha_i - \alpha_j)^2 + \alpha_k^2 \ge 0 \\
(\alpha_i + \alpha_j)^2 + \alpha_k^2 \ge \tfrac{1}{2},
\end{align}
where $i, j, k$ are all different. This and results of Eqs.(\ref{d0bounds},\ref{d3Equality}) gives
\begin{align}
\max \left(0,\, \frac{1-2 D_0}{4 D_0}\right) \le \tilde{D}_i \le \frac{1}{2}\\
\frac{1}{3} \le  D_0 \le 1\\
\tilde{D}_{p-1} + \tilde{D}_{q-p} + \tilde{D}_{q-1} = \frac{1- D_0}{2 D_0}.
\label{eq:k3GeneralBounds}
\end{align}
The first line defines a cube in $\tilde{D}_i$ space and the last two a family of planes that cut through that space. We show that within the cube the Hessian is always positive.

The function $R_3$, and therefore the Hessian, depends on the indices of the $\tilde{D}$ due to the $\sigma$ ``energy matching'' term in the triple sum. There are several triplets that could enter:
\begin{align}
& D_{p-1}\;D_{q-p}\;D_{q-1} \\
& \qquad \text{always contributes}\nonumber\\
& D_{p-1}\;D_{p-1}\;D_{q-1} \text{ or } D_{q-p}\;D_{q-p}\;D_{q-1} \\
& \qquad \text{are ruled out by the condition $p-1 \neq q-p$} \nonumber \\
& D_{p-1}\;D_{p-1}\;D_{q-p} \\
& \qquad \text{if and only if $q=3p - 2$} \nonumber\\
& D_{q-p}\;D_{q-p}\;D_{p-1} \\
& \qquad \text{if and only if $q=\tfrac{1}{2}(3p - 1)$} \nonumber
\end{align}
The first case is the generic one. The second case happens if the energy differences are equal, which we explicitly rule out. The third case happens if the populated levels are $(1, 2, 4),\, (1, 3, 7),\, ...$ where the energy difference is in the ratio $1:2$. The third case requires the populated levels to be $(1, 3, 4),\, (1, 5, 7),\, ...$ where the energy difference has the ratio $2:1$. This is therefore identical to the previous case under the Hamiltonian mapping $H\to-H$, which clearly leaves the interference pattern unchanged.

There are thus 2 different cases to consider. The Hessian in the first case is 
\begin{align}
\left(\begin{array}{c c c}
1 & \tilde{D}_{q-1} & \tilde{D}_{q-p} \\
\tilde{D}_{q-1} & 1 & \tilde{D}_{p-1} \\
\tilde{D}_{q-p} & \tilde{D}_{p-1} & 1
\end{array}\right).
\end{align}
This is positive definite as, from \eqref{eq:k3GeneralBounds} all principle minors of the matrix are themselves positive definite in the cubic region of interest \cite{Horn1985}. The second case has the Hessian
\begin{align}
\left(\begin{array}{c c c}
1 +\tilde{D}_{q-p} & \tilde{D}_{q-1} + \tilde{D}_{p-1}& \tilde{D}_{q-p} \\
\tilde{D}_{q-1} + \tilde{D}_{p-1} & 1 & \tilde{D}_{p-1} \\
\tilde{D}_{q-p} & \tilde{D}_{p-1} & 1
\end{array}\right),
\end{align}
which is also positive everywhere, except potentially at some of the vertices of the cube.

The vertices can be found in much the same way as before, except that the boundaries are now symmetric between the $\tilde{D}_i$. We therefore give them as triplets where all permutations need to be considered separately for evaluating $R_3$.
\begin{align}
\begin{array}{c | c c c | c c}
 D_0 & \tilde{D}_i & \tilde{D}_j & \tilde{D}_k & \max R_3 (\text{generic}) & \max R_3 (1:2 \text{ ratio})\\
\hline
\frac{1}{2} <  D_0 \le 1 & 0 & 0 & \frac{1- D_0}{2 D_0} & 1.25 & 1.25\\
\hline
\frac{1}{3} \le  D_0 \le \frac{1}{2} & \frac{1-2 D_0}{4 D_0} & \frac{1-2 D_0}{4 D_0} & \frac{1}{2} & 1.27 & 1.33\\
\end{array}
\end{align}

Importantly, these values are all lower than for the case of a 3-level system in adjacent energy levels. Therefore, the previous result we had is very significantly strengthened: if $R_3$ is larger than $1.86$ then we know that the state is not 3-coherent for any Hamiltonian.

%%%%%%%%%%
\subsection*{$k=d=4$}
To highlight that this algorithmic way of calculating the threshold values can be extended to high dimentions, we demonstrate it for the case of $4$-coherent states. In order to reduce the number of cases to consider, we limit ourselves to states where the $4$ populated levels are all adjacent basis states of an harmonic Hamiltonian. For this case, the variables are 
\begin{align*}
D_0 &= \alpha_1^2 + \alpha_2^2 + \alpha_3^2 + \alpha_4^2\\
D_1 &= \alpha_1 \alpha_2 + \alpha_2 \alpha_3 + \alpha_3 \alpha_4\\
D_2 &= \alpha_1 \alpha_3 + \alpha_2 \alpha_4 \\
D_3 &= \alpha_1 \alpha_4.
\end{align*}
Eqs.(\ref{d0bounds}, \ref{d3Equality}, \ref{Dibound}) hold as before. Other bounds can be obtained in a similar way to before by considering sums of squares. These are firstly
\begin{align}
(\alpha_1 - \alpha_3)^2 + (\alpha_2 - \alpha_4)^2 \ge 0 \nonumber \\
1 - 2 \tilde{D}_2 \ge 0,
\end{align}
and
\begin{align}
(\alpha_1 + \alpha_3)^2 + (\alpha_2 + \alpha_4)^2 \ge \tfrac{1}{2} \nonumber \\
 D_0(1 + 2 \tilde{D}_2) \ge \tfrac{1}{2}.
\end{align}
Similarly there is
\begin{align}
(\alpha_1 - \alpha_4)^2 + \alpha_2^2 + \alpha_3^2 \ge 0 \nonumber \\
1 - 2 \tilde{D}_3 \ge 0,
\end{align}
and
\begin{align}
(\alpha_1 + \alpha_4)^2 + \alpha_2^2 + \alpha_3^2 \ge \tfrac{1}{3} \nonumber \\
 D_0(1+ 2 \tilde{D}_3) \ge \tfrac{1}{3}.
\end{align}
Finally 
\begin{align}
(\alpha_1 - \alpha_2 + \alpha_3 - \alpha_4)^2 \ge 0 \nonumber\\
1 - 2 \tilde{D}_1 + 2 \tilde{D}_2 - 2 \tilde{D}_3 \ge 0.
\label{d43Body}
\end{align}

Rougher versions of these can be obtained by eliminating $ D_0$ by taking the `worst case' approach, providing the simple inequalities
\begin{align}
\tilde{D}_1 \le 1 &\qquad \tilde{D}_1 + \tilde{D}_3 \le 1\nonumber\\
\tilde{D}_2 \le \tfrac{1}{2} &\qquad \tilde{D}_3 \le \tfrac{1}{2},
\label{d4simple}
\end{align}
which will be useful in proving the positivity of the Hessian. The Hessian is given by
\begin{align}
\left(\begin{array}{c c c}
1 + D_2 & D_1 + D_3 & D_2 \\
D_1 + D_3 & 1 & D_1 \\
D_2 & D_1 & 1
\end{array}\right).
\end{align}
The easiest way to prove positivity is, as before, to show that each of the principle minors is itself positive definite in \cite{Horn1985}, which is straightforward to compute in the region of interest defined by the inequalities \eqref{d4simple}. As before, it remains to find the vertices as a function of $\tilde{D}_2$. This is a harder problem than before, which is most easily tackled by rewriting the tighter inequalities as
\begin{align}
&\max \left(0, \frac{1-2 D_0}{4 D_0}\right) \le \tilde{D}_2 \le \frac{1}{2}\\
&\max \left(0, \frac{1-3 D_0}{6 D_0}\right) \le \tilde{D}_3 \le \frac{1}{2}\\
&0 \le 1 - 2 \tilde{D}_1 + 2 \tilde{D}_2 - 2 \tilde{D}_3\\
&0 \le \tilde{D}_1 = \frac{1- D_0}{2  D_0} - \tilde{D}_2 - \tilde{D}_3 \le 1
\end{align}
The first two describe a surface in the $\tilde{D}_2 - \tilde{D}_3$, with the coordinates of the vertices depending on the value of $ D_0$. The third equation states how this protrudes in the $\tilde{D}_1$ direction. The fourth describes a plane that cuts this volume, and imposes additional physical constraints. The way to solve this is therefore, for a given range of $ D_0$, to find the vertices in the $\tilde{D}_2 - \tilde{D}_3$ plane (a maximum of 4), find the corresponding value of $\tilde{D}_1$ and check if any additional constraints on $ D_0$ arise. The results are summarised below.
\begin{align}
\begin{array}{c | c c c | c}
 D_0 & \tilde{D}_1 & \tilde{D}_2 & \tilde{D}_3 & \max R_3\\
\hline
\frac{1}{2} <  D_0 \le 1 & \frac{1 -  D_0}{4 D_0} & 0 & 0 & 1\\
\hline
 & \frac{1 - 2  D_0}{2 D_0} & \frac{1}{2} & 0 & 1.58\\
\frac{1}{3} <  D_0 \le \frac{1}{2} & \frac{1 - 2  D_0}{4 D_0} & \frac{1 - 2  D_0}{4 D_0} & \frac{1}{2} & 1.25\\
 & \frac{1}{4 D_0} & \frac{1 - 2  D_0}{4 D_0} & 0 & 1.86\\
\hline
 & \frac{1 - 3  D_0}{2 D_0} & \frac{1}{2} & \frac{1}{2} & 1.33\\
\frac{1}{4} \le  D_0 \le \frac{1}{3} & \frac{2 - 3  D_0}{6 D_0} & \frac{1}{2} & \frac{1 - 3  D_0}{6 D_0} & 2.44\\
 & \frac{1 - 2  D_0}{4 D_0} & \frac{1 - 2  D_0}{4 D_0} & \frac{1}{2} & 1.93
\end{array}
\end{align}

We see that the plane can intersect the volume at a single point (one vertex), in a plane (three vertices) or, in the case that $ D_0 = \tfrac{1}{4}$ in a line (two vertices). It is to be expected that this geometry becomes far more complicated in higher dimensions. This sort of analyses ought to generalise, but doing so is probably difficult. Nevertheless, from the table we can conclude that if $R_3$ is larger than $2.44$ we can certify that the state is not a 4-coherent state lying in adjacent energy levels of an SHO. For comparison, the perfectly balanced state gives $2.26$ and the largest value we could fine numerically was $2.32$. The largest value found for a 5-coherent state (that we want to distinguish from) was $2.88$.\\

%%%%%%%%%%%%%%%
\section{Derivation of $R_3(\ket{W_k}, \ket{W_k})$}
\label{sec:StraightLineR3}
We seek an exact analytical expression for the value of the certifier $R_3$ for the maximally coherent state, $R_3(\ket{W_k}, \ket{W_k})$. For the Hamiltonian in Eq. (\ref{eq:BasicHamiltonian}), the certifier can be rewritten as
\begin{equation}
	R_3(\ket{W_k}, \ket{W_k}) = \frac{1}{k} + \frac{12}{k^3} \frac{1}{T} \int_0^T \left( \sum_{i < j} \cos{(\omega_{i,j}t)} \right)^2 + \frac{8}{k^4} \frac{1}{T} \int_0^T \left( \sum_{i < j} \cos{(\omega_{i,j}t)} \right)^3, \label{eq:combinatorix}
\end{equation}
for energy level differences $\omega_{i,j} = |i-j|$.

The first and second integral involve products of two and three cosinusoidal terms respectively. Using the trigonometric identity for terms of frequencies $\alpha > \beta > \gamma \geq 0$,
\begin{equation}
	\cos{(\alpha)}\cos{(\beta)}\cos{(\gamma)} = \frac{1}{4} \Bigl[ \cos{(\alpha + \beta + \gamma)} + \cos{(-\alpha + \beta + \gamma)} + \cos{(\alpha - \beta + \gamma)} + \cos{(\alpha + \beta - \gamma)} \Bigr],
\end{equation}
these products are reduced into linear terms. We need to find those that survive and calculate the integrals for them. The condition $\alpha = \beta + \gamma$ is equivalent to the statement that at least one, and in fact exactly one, of the linearised terms survives. In other words, the largest energy level spacing must be equal to the sum of the two smaller ones. Once the conditions for non-vanishing terms in the products of cosines have been identified, it is a matter of counting the number of combinations $A$ and $B$ of energy levels that obey these conditions and survive in the first and second integral in Eq. (\ref{eq:combinatorix}) respectively, leading to:
\begin{equation}
	\label{eq:prelim_maxcoh}
	R_3 = \frac{1}{k} + \frac{6}{k^3} A + \frac{2}{k^4} B
\end{equation}

Calculating $A$ is simple, since in this case $\gamma = 0$ and the non-vanishing terms are the ones with identical cosines multiplied together. Therefore, summing over all different values of $\omega_{i,j}$ gives
\begin{equation}
	\label{eq:combinationsa}
	A = \sum\limits_{n=1}^{k-1} n^2 = \frac{k(k-1)(2k-1)}{6}.
\end{equation}

Calculating $B$ requires that cosine terms multiplied together satisfy that the largest frequency equals to the sum of the smaller ones. Let us label the largest frequency by $\omega_{i,i+\alpha}$, then it has multiplicity $(k-\alpha)$ and there are $S_{\alpha}$ ways that two frequencies can sum up to $\omega_{i,i+\alpha}$. Now, we seek all frequencies $\omega_{j_1,j_1+\beta}$ and $\omega_{j_2,j_2+\gamma}$ of multiplicities $(k-\beta)$ and $(k-\gamma)$ respectively, for which $\omega_{i,i+\alpha} = \omega_{j_1,j_1+\beta} + \omega_{j_2,j_2+\gamma}$, for all indices $i, j_1, j_2$. The last factor to consider is that the three cosines may be multiplied together in any order, so there is a combinatorial coefficient of $3!$ when three different frequencies are multiplied together and $\frac{3!}{2!}$ when the two shorter frequencies are the same, as in when $\beta = \gamma$, which can only happen for even $\alpha$. We now reach the expression
\begin{align}
	\label{eq:fa}
	S_\alpha &= 
	\begin{cases} 
      3!\sum\limits_{\substack{\beta+\gamma = \alpha \\ \beta \neq \gamma}} (k-\beta)(k-\gamma) & (\alpha \text{ is odd})
      \\[5ex]
      3!\sum\limits_{\substack{\beta+\gamma = \alpha \\ \beta \neq \gamma}} (k-\beta)(k-\gamma) + \frac{3!}{2!} \left( k - \frac{\alpha}{2} \right)^2 & (\alpha \text{ is even})
   \end{cases}
   \\[3ex]
   &= 	\frac{1}{2} (\alpha - 1) (\alpha + \alpha^2 - 6\alpha k + 6k^2),
\end{align}
for any $0 < \alpha < k$. Finally, summing over all allowed energy level differences,
\begin{equation}
	\label{eq:combinationsa}
	B = \sum\limits_{\alpha=1}^{k-1} (k-\alpha)S_{\alpha} = \frac{1}{40} k(k-1)(k-2)(2-7k+11k^2).
\end{equation}
Substituting $A$ and $B$ in Eq. (\ref{eq:prelim_maxcoh}), we get the desired sequence
\begin{equation}
	\label{eq:ratios_maxcoh}
	R_3 \left( \ket{W_k}\bra{W_k}, \ket{\chi_0} \right) = \frac{4 + 5k^2 + 11k^4}{20k^3}.
\end{equation}

%%%%%%%%%%%%%%%
\section{Derivation of decoherence theoretical and pattern thresholds}
\label{sec:decoh_thr}
We first derive the theoretical threshold of coherence for the Werner-like state $\rho_W$ of \eqref{eq:werner} and then prove that an interference pattern gives a threshold equal to the theoretical, under optimal measurement.

We observe that $\rho_W \in C_k$ is fully symmetric under permutations of basis states as well as that all the off-diagonal elements are $\frac{1-\lambda}{k}$, resulting in
\begin{equation}
	\label{eq:werner_norm}
	C_{\ell_1}(\rho_W) = (k-1)(1-\lambda),
\end{equation}
where $C_{\ell_1}(\rho) \coloneqq \sum\limits_{i \neq j} |\rho_{ij}|$ is the $\ell_1$-norm as studied by Bera \textit{et al}~\cite{ref:Bera}.
These two properties define a Werner-like state. 

In general, the $\ell_1$ norm of a $q$-coherent state is bounded from above. The bound is obtained when the system state is pure since $C_{\ell_1}$ is a convex measure~\cite{ref:Bera}. Let $\rho = \ket{\alpha}\bra{\alpha} \in C_q$ for a state $\ket{\alpha}$ defined in the reference basis, so that $\rho_{ij} = \alpha_i \alpha_j^* = \alpha_i^* \alpha_j$ and $\tr{\rho} = \sum_{i=1}^q |\rho_{ii}| = 1$.
\begin{align}
	(q-1) - C_{\ell_1}(\rho) = &(q-1) \sum_{i=1}^q |\rho_{ii}| - 2\sum\limits_{i < j} |\rho_{ij}| \\
	= &\sum\limits_{i < j} (|\alpha_i| - |\alpha_j|)^2 \geq 0.
\end{align}
This means that the coherence of the system is bounded above,
\begin{equation}
	\label{eq:cl1_bound}
	C_{\ell_1}\left(\rho\right) \leq q-1,
\end{equation}
with equality obtained when $\forall i,j, |\alpha_i| = |\alpha_j|$ in the reference basis, so that $\ket{\alpha}$ is the maximally $q$-coherent state.

Using Eqs.(\ref{eq:werner_norm}, \ref{eq:cl1_bound}), we obtain for the Werner-like states in $C_q$
\begin{align}
	\label{eq:werner_bound}
	\lambda &\geq \frac{k-q}{k-1} \\
	\therefore \lambda_{\text{dec}}(q) &= \frac{k-q}{k-1}, \quad 1 \leq q \leq k. 
\end{align}

Now projecting with the optimal measurement $\ket{W_q}$ we get
\begin{align}
	p(t)&= \matel{W_q}{\rho_W}{W_q} \\
	&\leq \frac{1}{k} + \frac{2}{k}\sum_{i < j} |\rho_{ij}| = \frac{1}{k} + \frac{1}{k}C_{\ell_1}\left(\rho\right)\\
	&\leq \frac{1}{k} + \frac{q-1}{k} = \frac{q}{k}
\end{align}

Therefore, a pattern with a maximum higher than this boundary value, $\frac{q}{k}$, cannot be decomposed into patterns arising from states of $q$-coherence or lower. We get the threshold value $\lambda_{\text{patt}}(q)$ at which the interference pattern can no longer distinguish consecutive coherence levels, by bounding the interference pattern produced from the Werner-like state by the probability maximum, so that
\begin{align}
	\frac{q}{k} &\geq \matel{W_q}{\rho_W}{W_q} = 1 - \lambda + \frac{\lambda}{k} \nonumber \\
	\Rightarrow \lambda &\geq \frac{k-q}{k-1} \nonumber \\
	\therefore \lambda_{\text{patt}}(q) &= \frac{k-q}{k-1}, \quad 1 \leq q \leq k, 
\end{align}
which coincides with $\lambda_{\text{dec}}(q)$.

\end{document}